\documentclass[%
reprint,
 amsmath,amssymb,
 aps,
showpacs,showkeys,]{revtex4-1}

\usepackage{graphicx}
\usepackage{dcolumn}
\usepackage{bm}

\begin{document}


\title{Energy-independent complex single-$P$-waves $NN$ potential\\
from Marchenko equation}

\author{N. A. Khokhlov}
\email{nikolakhokhlov@yandex.ru; nakhokhlov@gmail.com}
\affiliation{%
Peter the Great St. Petersburg Polytechnic University, RU-195251 St. Petersburg, Russia. 
}%
\date{\today}

\begin{abstract}
 We extend our previous results of solving the inverse problem of quantum scattering theory (Marchenko theory, fixed-$l$ inversion). In particular, 
we apply an isosceles triangular-pulse function set for the Marchenko equation input kernel expansion in a separable form. The separable form allows a reduction of the Marchenko equation to a system of linear equations for the output kernel expansion coefficients.  We show that in the general case of a  single partial wave, a linear expression of the input kernel is obtained in terms of the Fourier series coefficients of $q^{1-m}(1-S(q))$ functions in the finite range of the momentum $0\leq q\leq\pi/h$ [$S(q)$ is the scattering matrix, $l$ is the angular orbital momentum, $m=0,1,\dots,2l$]. Thus, we show that the partial  $S$--matrix on the finite interval  determines a  potential function with $h$-step accuracy. The calculated partial potentials describe a partial $S$--matrix with the required accuracy. The partial $S$--matrix is unitary below the threshold of inelasticity and non--unitary (absorptive) above the threshold. 
We developed a  procedure and applied it to partial-wave analysis (PWA) data of $NN$ elastic scattering  up to 3 GeV. We show that energy-independent complex partial potentials describe these data for single $P$-waves. 
\end{abstract}
\maketitle

\section{\label{sec:intro}INTRODUCTION}
Extracting the interparticle interaction potential from scattering data is a fundamental problem in nuclear physics. The various inverse problem of quantum scattering (IP) methods are used for such extraction. 
Fitting parameters of the phenomenological potential can solve this problem. Such an adjustment is easily implemented in the case of a small number and low accuracy of the known experimental data. An accurate description of a large number of experimental data requires exact methods of IP solving. The development of such precise and unambiguous methods remains a fundamental challenge \cite{Sparenberg1997,Sparenberg2004,Kukulin2004,Pupasov2011,Mack2012}. The ill-posedness of the problem significantly complicates
its numerical solution.

The fixed-$l$ IP considered here is usually solved within 
Marchenko, Krein, and Gelfand-Levitan theories \cite{Gelfand, Agranovich1963, Marchenko1977,Blazek1966, Krein1955, Levitan1984, Newton, Chadan}.
In these approaches, the IP is reduced to solving Fredhölm integral equations of the second kind. H.V.~von~Gevamb and H.~Kohlhoff  successfully applied Marchenko and  Gelfand-Levitan theories to extract $NN$ partial potentials from PWA data \cite{Geramb1994, Kohlhoff1994}.  They used the PWA data up to the inelastic threshold ($E_{\text{lab}}\approx~280$~MeV) and approximated the corresponding partial $S$--matrices (spectral densities) by rational fraction expansions (Padé approximants). In this case, the input kernels of the integral equations are represented as finite separable series of the Riccati-Hankel functions products (separable kernels). A Fredhölm integral equation of the second kind with a separable kernel is solved analytically. The partial potentials, in this case, are also expressed through Riccati-Hankel functions (Bargman-type potentials). A similar approach in frames of the Marchenko theory was used later to extract optical model $NN$ partial potentials from elastic $NN$ scattering PWA data up to 3~GeV \cite{Khokhlov2006, Khokhlov2007}. This method for solving the inverse problem includes four numerical procedures. The first procedure is the PWA, the second procedure is the approximation of the $S$-matrix, the third procedure is the solution of the integral equation, and the fourth procedure is the differentiation of the output kernel to calculate the potential. Errors of each step are accumulated, the estimation of the error in the calculation of the potential is not an easy task.
For $l\ge 1$, Marchenko  inversion is unstable for $r<1$~fm, and Gelfand-Levitan is unstable for $r>8$~fm \cite{Geramb1994, Kohlhoff1994}. Description of PWA data within errors for energies up to 3~GeV requires the use of high-order Padé approximants \cite{Khokhlov2006, Khokhlov2007}. We found that an increase in the accuracy of the approximation and, accordingly, an increase in the order of the Padé approximant can lead to significant changes in the potential. The Padé approximant for such a problem is not the best choice since approximants of different orders giving close $S$-matrix values at the PWA points can differ significantly between these points. Thus, the convergence of methods using rational fraction expansions of $S$-matrix (spectral density) is not apparent. 
 
This paper generalizes the algebraic method \cite{MyAlg1,MyAlg2} for solving the IP. 
We expand the Marchenko equation input kernel into a separable series in an isosceles triangular-pulse function set. 
 We obtained a linear expression of the expansion coefficients in terms of the Fourier series coefficients of $q^{1-m}(1-S(q))$ ($m=0,1,\dots,2l$) functions on a finite range of the momentum $0\leq q\leq\pi/h$.
Thus, we solve the Marchenko equation with a separable kernel, which can be performed analytically like in
Refs.~\cite{Geramb1994, Kohlhoff1994, Khokhlov2006, Khokhlov2007}. 
 Theory of the Fourier's series substantiates convergence of the procedure with decreasing step $h$.

$NN$ potentials used as an input for  (semi)microscopic construction of nuclear optical potentials (OPs)
are usually real and only describe the $NN$ PWA data below the inelastic threshold \cite{Burrows2020,Vorabbi2018,Vorabbi2017,Guo2017,Vorabbi2016}.  
However, in the absence of a microscopic theory, complex partial $NN$ potentials are required to describe nuclear reactions with energies of the $NN$ relative motion above the threshold \cite{Khokhlov2007}. 
In existing models, real partial potentials are modified by energy-dependent imaginary terms which are zero below the threshold  
\cite{Khokhlov2006,Funk2001}. The partial wave function is real below the threshold to within an $r$-independent factor in such models. 
For realistic optical potentials, such a restriction is unreasonable.
Indeed, the optical model potential (pseudopotential) definition does not guarantee that it will be Hermitian \cite{Taylor,Goldberger}. One can only assert that Hamiltonian eigenfunctions corresponding to eigenenergies below the threshold satisfy the Hermitianity condition. 
However, the Hermitian condition 
\begin{equation}
	\int \psi^{*}(V^{+}-V)\psi dx =0 \label{Herm}
\end{equation}
implies $V^{+}-V=0$ only if Eq.~\ref{Herm} holds for an arbitrary function $\psi$, which is not true for optical potentials.  
Using the phase-equivalent Krein transformations, one can obtain an energy-independent complex potential giving a unitary matrix \cite{Agranovich1963}. The use of optical potentials limited by the condition $V^{+}-V=0$ below the inelasticity threshold is a physically unreasonable limitation. Previously, it was shown that the Marchenko theory applies to the description of elastic $nD$ scattering from zero energy up to energies well above the threshold. 
In this case, Marchenko theory produces energy-independent complex partial $nD$ potentials \cite{Papastylianos1990, Alt1994}. 
They used a rational parametrization, the same as in Refs.~\cite{Geramb1994, Kohlhoff1994, Khokhlov2006, Khokhlov2007} for the unitary $S$--matrix.
The small value of the threshold (the deuteron's binding energy $E_{\text{c.m.}}\approx 2.226$~MeV) does not allow us to judge the applicability of the Marchenko theory in the general case. 
We analyzed the Marchenko theory \cite{Agranovich1963, Marchenko1977} ($l=0$), \cite{Blazek1966} ($l>0$)   and found that most of the theory applies to unitary and to non--unitary $S$--matrices. 
Our algebraic form of the Marchenko equation allows calculating the energy-independent complex local partial potential corresponding to a partly unitary and partly non--unitary $S$--matrix. 
We previously applied this approach to analyze $^{1}S_{0}$ data of elastic $NN$ scattering PWA \cite{MyAlg2}.
Extending this  investigation, we analyze single $P$-waves $NN$ PWA data (up to $E_{\text{lab}} \approx 3$~GeV) and show that these data are described by energy-independent complex partial potentials. 
The reconstructed partial $S$ and  $P$-waves potentials constitute energy-independent soft core OP that describes 
elastic $NN$ scattering PWA data up to 3~GeV. 

\section{Marchenko equation in an algebraic form}
The radial Schrödinger equation is
\begin{equation}
	\label{f1}
	\left(\frac{d^{2}}{d r^{2}}-\frac{l(l+1)}{r^{2}}-V(r)+q^{2}\right) \psi(r, q)=0.
\end{equation}
The Marchenko equation  \cite{Agranovich1963, Marchenko1977} is a Fredhölm integral equation of the second kind:
\begin{equation}
	\label{f3}	F(x, y)+L(x, y)+\int_{x}^{+\infty} L(x, t) F(t, y) d t=0
\end{equation}
The kernel function is defined by the following expression
\begin{multline}
	F(x, y)=\frac{1}{2 \pi} \int_{-\infty}^{+\infty} h_{l}^{+}(q x)[1-S(q)] h_{l}^{+}(q y) d q \\	
	+\sum_{j=1}^{n_{b}} h_{l}^{+}\left(\tilde{q}_{j} x\right) M_{j}^{2} h_{l}^{+}\left(\tilde{q}_{j} y\right)\\
	=\frac{1}{2 \pi} \int_{-\infty}^{+\infty} h_{l}^{+}(q x) Y(q) h_{l}^{+}(q y) d q
	\label{f4}	
\end{multline}
where $h_{l}^{+}(z)$ is the Riccati-Hankel function, and
\begin{equation}
	\label{f5}	Y(q)=\left[1-S(q)-i \sum_{j=1}^{n_{b}} M_{j}^{2}\left(q-\tilde{q}_{j}\right)^{-1}\right],
\end{equation}
Experimental data entering the kernel are
\begin{equation}
	\label{f2}
	\left\{S(q),(0<q<\infty), \tilde{q}_{j}, M_{j}, j=1, \ldots, n\right\},
\end{equation}
where $S(q)=e^{2 \imath \delta(q)}$  is a scattering matrix dependent on the momentum $q$. The $S$--matrix defines asymptotic behavior at  $r \rightarrow+\infty$ of regular at $r=0$  solutions of Eq.~(\ref{f1}) for $q \geq 0 ;\ \tilde{q}_{j}^{2}=E_{j} \leq 0, E_{j}$  is $j$--th bound state energy ($-\imath \tilde{q}_{j} \geq 0$); $M_{j}$  is $j$-th bound state asymptotic constant.

The potential function of Eq.~(\ref{f1}) is obtained from the solution of Eq.~\ref{f3}
\begin{equation}
	V(r)=-2 \frac{d L(r, r)}{d r} \label{f6}
\end{equation}
Many methods for solving  Fredhölm integral equations use a series expansion of the equation kernel. \cite{eprint7,eprint8,eprint9,eprint10,eprint12,eprint13,eprint14}. We also use this approach. 

We introduce auxiliary functions: 
\begin{equation}
	\label{f11b}	F_{m}(z)=\frac{1}{2 \pi} \int_{-\infty}^{+\infty}\frac{e^{\imath qz} Y(q)dq}{q^{m}},
\end{equation}
then
\begin{equation}
	\label{Fm_der}	\frac{d^{k}F_{m}(z)}{dz^{k}}=\imath^{k} F_{m-k}(z),\ (k=1,2,\dots, m).
\end{equation}

\begin{figure}[htb]
	\centerline{\includegraphics[width=0.46\textwidth]{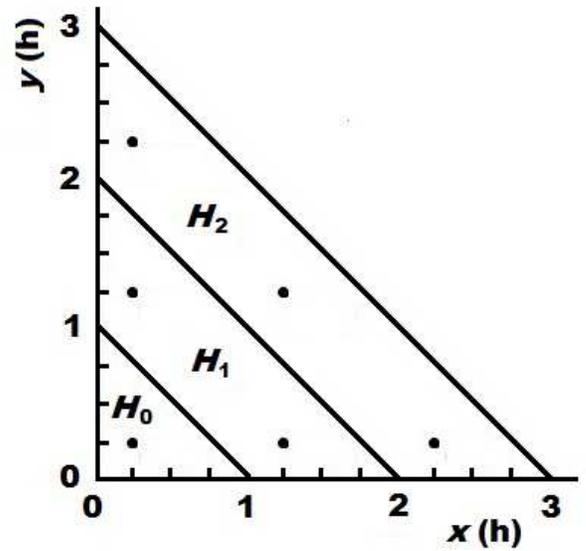}}
	\caption{\label{fig:basis1} The basis set $H_{n}\equiv H_{n}(x+y)$ (Eq.~(\ref{rec_set}) is shown as trapezoid (triangle for $n=0$) regions where  $H_{n}(x+y)=1$, and elsewhere $H_{n}(x+y)=0$. The regions are bounded by lines $x=0$, $y=0$, and $x+y=h(n-1)$.
		The basis set $\Delta_{i}(x)\Delta_{j}(y)$ (Eq.~(\ref{tr_set}))  is shown as projections (points) of the corresponding regular square pyramids apexes on the $xy$--plane.
		$\Delta_{i}(x)\Delta_{j}(y)=1$ at $x=(0.25+i)h,\ y=(0.25+j)h$ (apex of the $ij$--pyramid).
		The pyramids bases are $(2h\times 2h)$ squares on the $xy$--plain with sides parallel to the $x$ and $y$ axes. $\Delta_{i}(x)\Delta_{j}(y)=0$ on sides of the corresponding squares (and outside them).}
\end{figure}

\begin{widetext}
We use transformations 
\begin{equation}\label{trKrein1}
	\hat{K}_{z,l}f(z)= z^{l+1}\left(-\frac{1}{z}\frac{d}{dz}\right)^{l}\left[z^{-1}f(z)\right] 
	\equiv (-1)^{l} {\sum_{n=0}^{l}\frac{(2l-n)!}{n!(l-n)!}(-2z)^{n-l}\frac{d^{n}f(z)}{dz^{n}} 	}.
\end{equation}
From \cite{Abramowitz} (Eqs.~(10.1.23)--(10.1.26)) we get
\begin{equation}
	\label{trKrein1h1}	
	\hat{K}_{z,l}e^{\pm \imath qz}=q^{l} h^{\pm}_{l}(qz).
\end{equation}
	and
	\begin{multline}
		\hat{K}_{y,l}	\hat{K}_{x,l}F_{2l}(x+y)
		= {\sum_{n_{1},n_{2}=0}^{l}\frac{(2l-n_{1})!}{n_{1}!(l-n_{1})!} \frac{(2l-n_{2})!}{n_{2}!(l-n_{2})!}
			(-2x)^{n_{1}-l}	(-2y)^{n_{2}-l}\imath^{n_{1}+n_{2}}F_{2l-n_{1}-n_{2}}(x+y)  	} \\
		=\frac{1}{2 \pi} \int_{-\infty}^{+\infty} h_{l}^{+}(q x) Y(q) h_{l}^{+}(q y) d q\equiv F(x, y). \label{FKrein}
	\end{multline}
\end{widetext}
%
Assuming the finite range $R$ of the potential function $V(r)$, we approximate $F_{m}(x+y)$ as follows:
\begin{eqnarray}
	F_{m}(x+y)\approx \sum_{k=-2 N}^{2 N} f_{m,k} H_{k}(x+y) \label{rec_apr}\\
	\approx \sum_{k, j=0}^{N} \Delta_{k}(x) f_{m,k+j} \Delta_{j}(y) \label{tr_apr} 
\end{eqnarray}
where $f_{m,k} \equiv F_{m}(k h)$, $h$ is some step, and $R = Nh$. 
The used basis sets are 
\begin{equation}
	\left. \begin{array}{l} H_{0}(x)=\left\{\begin{array}{lr}
			1 & \text{if }  0 \leq x \leq h, \\
			0 & \text{otherwise,}
		\end{array}\right.\\
		H_{n}(x)=H_{0}(x-h n).
	\end{array}\right\} \label{rec_set}
\end{equation}
\begin{equation}
	\left. \begin{array}{l}	\Delta_{0}(x)=\left\{\begin{array}{lr}
			1-|x-0.25| / h & \text{if }  |x-0.25| \leq h, \\
			0 & \text{otherwise,}
		\end{array}\right.\\
		\Delta_{n}(x)=\Delta_{0}(x-h n). 
	\end{array}\right\} \label{tr_set}
\end{equation}
We use bases set $\Delta_{i}(x)\Delta_{j}(y)$ shifted by the vector $(0.25h,0.25h)$ compared to the set used previously \cite{MyAlg1,MyAlg2}. 
The basis sets are illustrated in Fig.~\ref{fig:basis1}. 
Decreasing the step $h$, one can approach $F_{m}(x+y)$ arbitrarily close at all points with both sets. Coefficients $f_{m,k}$ are same for both approximations Eqs.~(\ref{rec_apr}), (\ref{tr_apr}). 

The Fourier transform of the basis set Eq.~\ref{rec_apr} 
\begin{equation}
	\tilde{\mathrm{H}}_{k}(q)=\int_{-\infty}^{\infty} \mathrm{H}_{k}(x) e^{-\imath q x} d x=\frac{\imath\left(e^{-\imath q h}-1\right)}{q e^{\imath q h k}}. \label{f13b}
\end{equation}
The Fourier transform of Eq.~(\ref{f11b}) yields
\begin{equation}
	\frac{Y(q)}{q^{m}}\approx \sum_{k=-2 N}^{2 N} f_{m,k} \tilde{\mathrm{H}}_{k}(q)=\sum_{k=-2 N}^{2 N} f_{m,k}  \frac{\imath\left(e^{-\imath q h}-1\right)}{q e^{\imath q h k}}. \label{f13c}
\end{equation}
We rearrange the last relationship
	\begin{multline}
		Y(q)/q^{m-1}= \imath \sum_{k=-2 N}^{2 N} f_{m,k} \left(e^{-\imath q h}-1\right)  e^{-\imath q h k}\\
		=\imath \sum_{k=-2 N+1}^{2 N}\left(f_{m,k-1}-f_{m,k}\right) e^{-\imath q h k}
		+\imath\left(-f_{m,-2 N}\right) e^{\imath q h 2 N}\\
		+\imath\left(f_{m,2 N}\right) e^{-\imath q h(2 N+1)}.
		\label{f13dn}	
	\end{multline}
	Thus, the left side of the expression is represented as a Fourier series on the interval $-\pi / h \leq q \leq \pi / h$. 
	\begin{equation}
		f_{m, k-1}  - f_{m,k}
		= -	\frac{ \imath h}{2\pi} \int_{-\pi / h}^{\pi / h}    Y(q)\frac{ e^{\imath q h  k}d q}{q^{m-1}} 	 \label{f14n}
	\end{equation}
for  $k=-2 N, \ldots, 2 N$. Recursive solving of the Eq.~(\ref{f14n}) from  $k=2 N+1$ ($f_{m,2 N+1}=0$) gives 
\begin{multline}
	f_{m,k} 
	  	=	\frac{h}{\pi} \int_{-\pi / h}^{\pi / h} \left( 
	  	\sum_{\nu = k+1}^{2N+1} e^{\imath q h  \nu}\right) 
	  \frac{ Y(q)d q}{q^{m-1}}\\
	=  -	\frac{\imath h}{2\pi} \int_{-\pi / h}^{\pi / h}
	\frac{e^{\imath q h (k+1)}\left(1-e^{\imath q h(2N-k+1)} \right)}{\left(1-e^{\imath q h}\right) {q^{m-1}} } Y(q)  
	d q.	 	
	\label{f14nn}
\end{multline}

The $F(x,y)$ is defined by $f_{m,k}$ $(m=0,1,\dots,2l)$, $k=0,1,\dots,2N$ from Eqs.~(\ref{FKrein}), and (\ref{tr_apr}) as 
\begin{equation}
	F(x, y) \approx  \sum_{k, j=0}^{N} \Delta_{k}(x) F_{k,j}\Delta_{j}(y), \label{f7nn}
\end{equation}
where 
\begin{widetext}
	\begin{eqnarray}
		F_{k,j} ={ {\sum_{n_{1},n_{2}=0}^{l}\frac{(2l-n_{1})!}{n_{1}!(l-n_{1})!} \frac{(2l-n_{2})!}{n_{2}!(l-n_{2})!}
				(-2(k+0.25)h)^{n_{1}-l}	(-2(j+0.25)h)^{n_{2}-l} \imath^{n_{1}+n_{2}} f_{2l-n_{1}-n_{2},k+j}  	}} \\
		=
		-	\frac{\imath h}{2\pi} \int_{-\pi / h}^{\pi / h}  
		h_{l}^{+}(q(k+0.25)h) \frac{e^{\imath q h}\left(1-e^{\imath q h(2N-k-j+1)} \right)}{1-e^{\imath q h}} Y(q)h_{l}^{+}\left(q(j+0.25)h\right)  qd q.	 
		\label{f14}
	\end{eqnarray}
\end{widetext}

Thus, the range of known scattering data defines the value of $h$   and, therefore, the inversion accuracy.

We solve the Eq.~(\ref{f3}) substituting
\begin{equation}
	\label{f9n}	L(x, y) \approx \sum_{j=0}^{N} P_{j}(x) \Delta_{j}(y)
\end{equation}
Substitution of Eqs.~(\ref{f7nn}) and (\ref{f9n}) into Eq.~(\ref{f3}) and linear independence of the basis functions give 
\begin{widetext}
	\begin{equation}
		\sum_{m=0}^{N}\left(\delta_{j\, m}+\sum_{n=0}^{N}\left[\int_{x}^{max((m+0.25)h,(n+0.25)h)} \Delta_{m}(t) \Delta_{n}(t) d t\right] F_{n,j}\right) P_{m}(x)	
		=-\sum_{k=0}^{N} \Delta_{k}(x) F_{k,j}
		\label{f10n}	
	\end{equation}
	We define
	\begin{equation}
		\zeta_{n\, m\, p}=\int_{(p+0.25) h}^{max((m+0.25)h,(n+0.25)h)} \Delta_{m}(t) \Delta_{n}(t) d t	
		=\frac{h}{6}\left(2 \delta_{n\, m}\left(\delta_{n\, p}+2 \eta_{n \geq p+1}\right)
		+\delta_{n\,(m-1)} \eta_{n \geq p}+\delta_{n\,(m+1)} \eta_{m \geq p}\right),
		\label{f10b}	
	\end{equation}
\end{widetext}
where $\delta_{k\, p}$ are the Kronecker symbols  $\delta_{k\, p}$, and
\begin{equation}
	\eta_{a}=
	\left\{\begin{array}{lr}
		1 & \text{if }  a \text{ is true}, \\
		0 & \text{otherwise,}
	\end{array}\right.
	\label{eta}	
\end{equation} 
Since  $\Delta_{k}(h p) \equiv \delta_{k\, p}$, we finally get a system of equations
\begin{equation}
	\label{f11n}	\sum_{m=0}^{N}\left(\delta_{j\, m}+\sum_{n=p}^{N} \zeta_{n\, m\, p} F_{n,j}\right) P_{p, m}=-F_{p,j},
\end{equation}
for $P_{k}(h (p+0.25)) \equiv P_{p, k}$ $(p,k = 0,\dots,N)$ ($j,p = 0,\dots ,N$).   

Solution of Eq.~(\ref{f11n}) gives  $P_{p, k}$. We calculate potential values at points $r = hp$ $(p = 0,\dots, N)$ from Eq.~(\ref{f6}) by some finite difference formula.  

\begin{figure}[htb]
	\centerline{\includegraphics[width=0.46\textwidth]{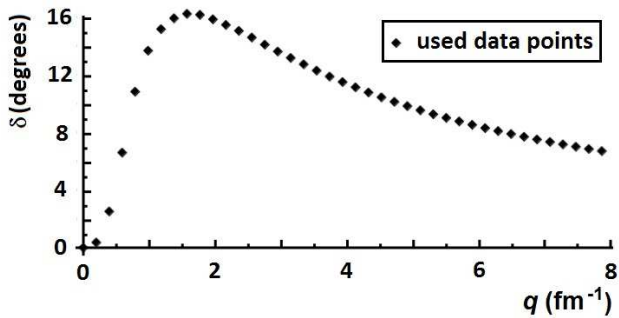}}
	\caption{\label{fig:expphase}Data used to reconstruct $V(r)=V_{0}\exp(-ar)$, where $V_{0}=-3\ fm^{-2}= -124.5\ MeV$, $a=1.5\ fm^{-1}$. Angular orbital momentum $l=1$. Units correspond to the $NN$ system.}
\end{figure}

\begin{figure}[htb]
	\centerline{\includegraphics[width=0.46\textwidth]{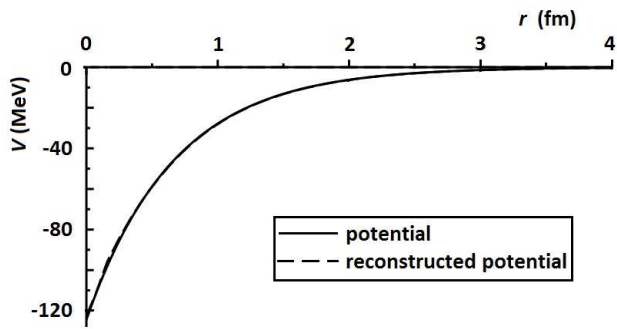}}
	\caption{\label{fig:exppot}Initial and reconstructed potentials:  $V(r)=V_{0}\exp(-ar)$, where $V_{0}=-3\ fm^{-2}= -124.5\ MeV$, $a=1.5\ fm^{-1}$. Units correspond to the $NN$ system.}
\end{figure}

We tested the developed approach by restoring the potential function  $V(r)=-3 \exp (-3 r / 2)$ from the corresponding scattering data.    Results are presented in Figs.~\ref{fig:expphase},~\ref{fig:exppot}, where  $h = 0.04$, $R = 4$. 
The input $S$--matrix was calculated at points shown in Fig.~\ref{fig:expphase} up to $q = 8$.  The $S$--matrix was interpolated by a quadratic spline in the range $0 < q < 8$. The $S$--matrix was approximated as asymptotic $S(q)\approx\exp (-2 i \alpha / q)$  for $q>8$, where  $\alpha$ was calculated at $q = 8$. 

\section{Energy-independent complex partial potentials}

Two-particle relativistic potential models can be represented in a non--relativistic form \cite{Keister1991}. Thus, the applicability of methods for solving the inverse problem is not limited only to nonrelativistic quantum mechanics. As we showed earlier \cite{MyAlg2}, modern data of the $NN$ partial-wave analysis up to $E_{\text{lab}}=3$~GeV can be described by the energy-independent complex partial potential for $^{1}S_{0}$ single wave at least.

After analyzing the Marchenko theory \cite{Agranovich1963,Marchenko1977}, including for the case of $l>0$ \cite{Blazek1966}, we found that Eqs.~(\ref{f1}, \ref{f4}) apply  to non--unitary $S$--matrices describing absorption. In this case, the absorbing partial $S$--matrix should be defined as  
\begin{equation}
 S(q)=\left\{\begin{array}{lll}
			S_{u}(q)+S_{n}(q) & \text { for } & q>0, \\
			S^{+}_{u}(-q)-S^{+}_{n}(-q) & \text { for } & q<0, 
		\end{array}\right.
		\label{OP_Smatrix}
\end{equation}
where superscript $+$ means hermitian conjugation. For $q>0$, we define
 \begin{equation}
 	S_{u}(q) = e^{2\imath \delta(q)},\ \
 		S_{n}(q) = -\sin^2(\rho(q))e^{2\imath \delta(q)},
 \end{equation}
where $\delta(q)$ and $\rho(q)$ are phase shift and inelasticity parameter correspondingly. 
With the $S$--matrix defined in this way, all Eqs.~(\ref{f14}--\ref{f11n}) remain valid and allow calculating local and energy-independent OP from an absorptive $S$--matrix and corresponding bound states' characteristics.

\section{Results and Conclusions}
\begin{figure}[h]
	\centerline{\includegraphics[width=0.5\textwidth]{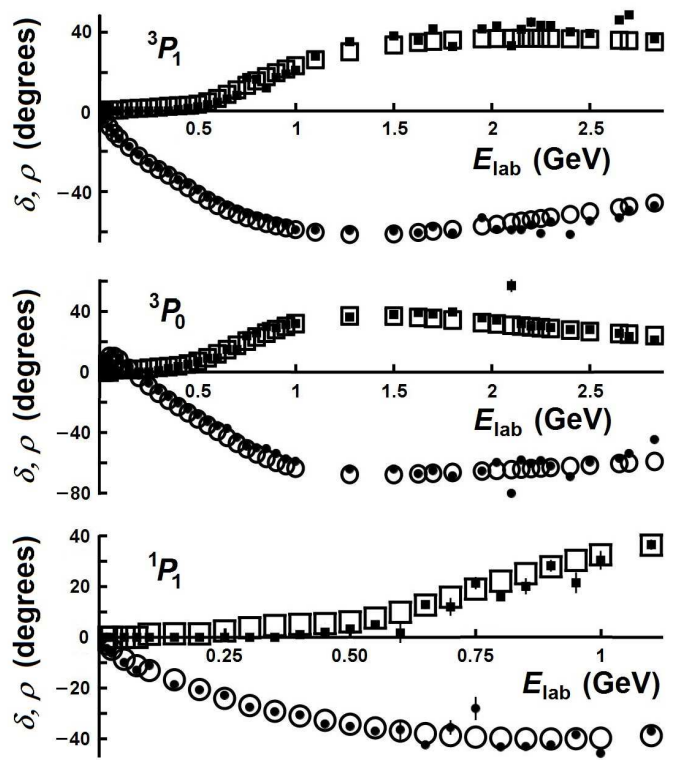}}
	\caption{\label{fig:Pdata} Data used to reconstruct $P$-wave partial potentials are  phase shifts $\delta$ (solid circles) and inelasticity parameters $\rho$ (solid squares) (from Ref~\cite{DataScat,cite}). Circles and squares stand for $\delta$ and $\rho$ correspondingly calculated from the reconstructed partial potentials. }
\end{figure}

Extending our previous results \cite{MyAlg2}, we analyzed modern single $P$-wave $NN$  phase shift  data  up to
3 GeV \cite{DataScat,cite}  (SW16, single-energy solutions,  Fig.~\ref{fig:Pdata}). 

We smoothed the phase shift and inelasticity parameter data   for $q>3$~\text{fm}$^{-1}$ by the following functions, correspondingly:
\begin{equation}
	\delta(q) \sim \sum_{k=3}^{8} A_{k}/q^{k},\ \
	\rho(q)\sim \sum_{k=1}^{9} A_{k}/q^{k},
	\label{asymp}
\end{equation}
where the coefficients were fitted by the least-squares method. The SW16 data and asymptotics (\ref{asymp}) above $q>3$~\text{fm}$^{-1}$ were used to calculate coefficients of Eqs.~(\ref{f14}) with $h=0.02$~fm corresponding to  $q_{\text{max}}\approx 157.08$~fm$^{-1}$.

\begin{figure}[h]
	\centerline{\includegraphics[width=0.5\textwidth]{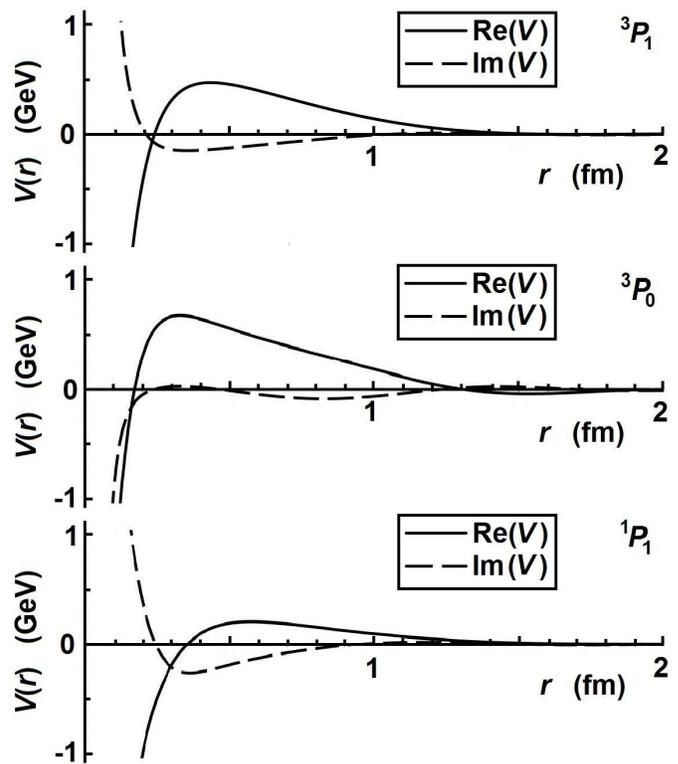}}
	\caption{\label{fig:Ppots} Real and imaginary parts of the reconstructed  partial potentials.}
\end{figure}

\begin{figure}[h]
	\centerline{\includegraphics[width=0.5\textwidth]{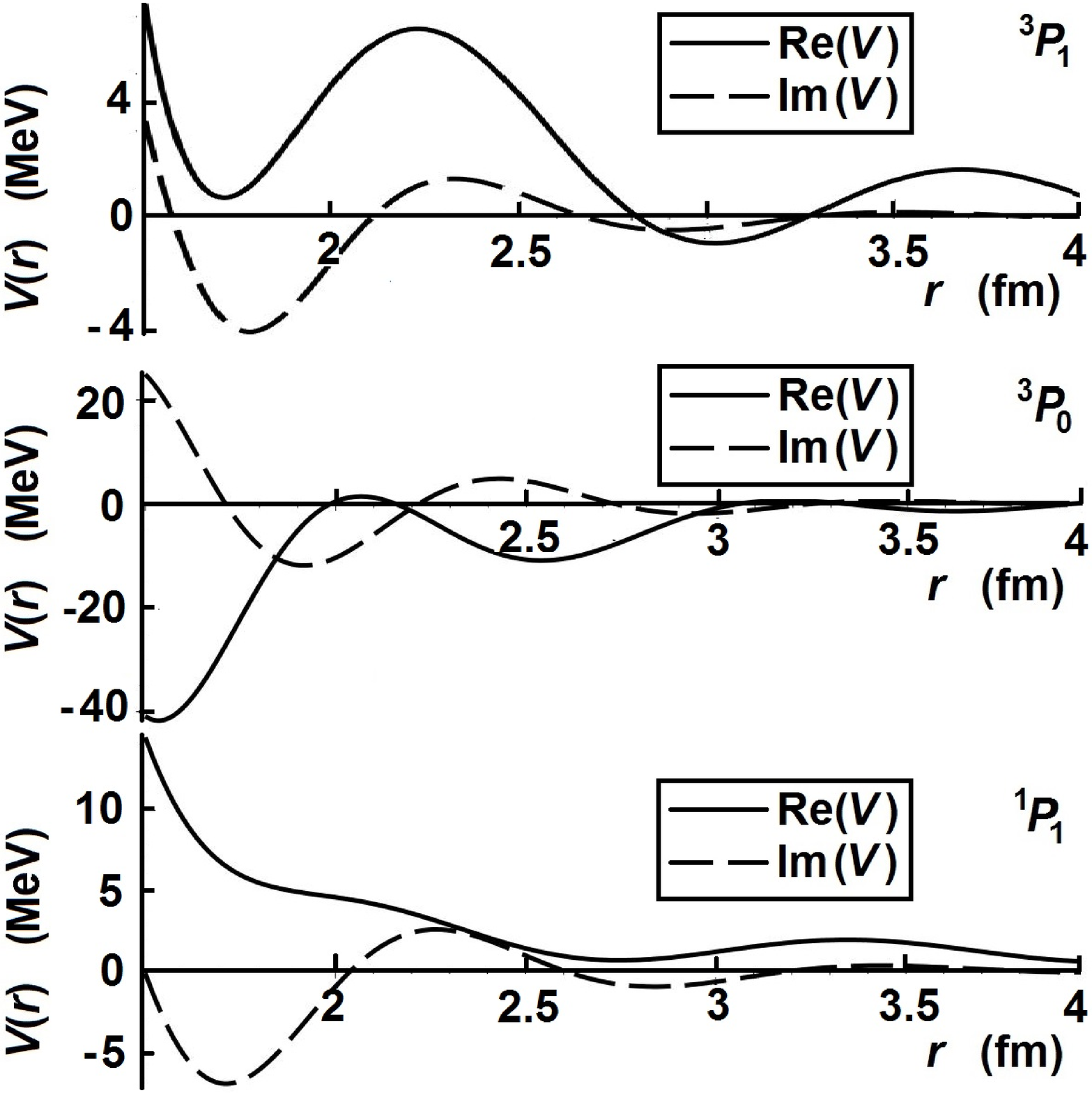}}
	\caption{\label{fig:PpotsA} Real and imaginary parts of the reconstructed  partial potentials.}
\end{figure}

The used  data are described by  energy-independent complex partial potentials (Figs.~\ref{fig:Ppots}, \ref{fig:PpotsA}). 
 Thus, we presented an IP solution algorithm for finite range potentials  (fixed-$l$, single partial wave inversion, in frames of the Marchenko theory).
%

Our method gives a geometric glimpse into the $NN$ interaction. 
As noted earlier by P.~Fernández-Soler and E.~Ruiz Arriola \cite{Fernandez-Soler2017}, "short-wavelength fluctuations/oscillations are inherent to the maximum energy or CM momentum being fixed for the phase shift".  
In this conclusion, they refer to the results of our calculations \cite{Khokhlov2006}. Then, we did not pay attention to this feature of the IP solution, and we considered the oscillations  an artifact of the used IP solution in which the potential is explicitly, algebraically expressible in terms of Hankel functions.  
Reconstructed   $^{1}S_{0}NN$ partial potential \cite{MyAlg2} shows that 
short-wavelength oscillations may indeed be a manifestation of a physical phenomenon. There are  also similar oscillations in the reconstructed $P$-wave partial potentials (Fig.~\ref{fig:PpotsA}).
 Microscopic calculations also predict such oscillations   \cite{Wendt2012}.

We show that IP solutions give optical model energy independent $NN$ partial potentials not only in the $^{1}S_{0}NN$ wave, but also in other single $NN$ partial waves.
The optical model partial $NN$ potentials with a repulsive core does not necessarily exhibit a strong energy dependence up to 3 GeV, as previously stated in \cite{Fernandez-Soler2017}. Instead of the soft repulsive core that was reconstructed in our model for the real part of the $^{1}S_{0}NN$ partial potential, the barriers at about 0.2-1.3~fm were reconstructed for the real parts of the $P$-waves partial potentials.

The reconstructed $P$-wave $NN$ complex potentials may be requested from the author in the Fortran code.

\end{document}